\documentclass[aps,pra,twocolumn,groupedaddress]{revtex4-1}

\usepackage{epsfig,amssymb,amsmath,amsfonts}
\usepackage{times}
\usepackage{graphicx}
\usepackage{epstopdf}

\def\>{\rangle}
\def\<{\langle}

\newcommand{\ignore}[1]{}

\def\bmat{\left[\begin{matrix}} 
\def\emat{\end{matrix}\right]}

\newcommand{\bra}[1]{\langle#1|}
\newcommand{\ket}[1]{{|}#1{\rangle}}

\begin{document}

\title{Approximation of real errors by Clifford channels and Pauli measurements}

\author{Mauricio Guti\'errez$^1$}
\author{Lukas Svec$^2$}
\author{Alexander Vargo$^3$}
\author{Kenneth R.\ Brown$^1$}
\email{ken.brown@chemistry.gatech.edu}
\affiliation{$^1$Schools of Chemistry and Biochemistry; Computational Science and Engineering; and Physics\\ 
Georgia Institute of Technology, Atlanta, GA 30332-0400}
\affiliation{$^2$Department of Physics, University of Washington, Seattle, WA 98195-2350}
\affiliation{$^3$Department of Mathematics and Statistics, Haverford College, Haverford, PA 19041-1336}

\date{\today}

\begin{abstract}

The Gottesman-Knill theorem allows for the efficient simulation of stabilizer-based quantum error-correction circuits. Errors in these circuits are commonly modeled as depolarizing channels by using Monte Carlo methods to insert Pauli gates randomly throughout the circuit. Although convenient, these channels are poor approximations of common, realistic channels like amplitude damping. Here we analyze a larger set of efficiently simulable error channels by allowing the random insertion of any one-qubit gate or measurement that can be efficiently simulated within the stabilizer formalism. Our new error channels are shown to be a viable method for accurately approximating real error channels.  

\end{abstract}


\maketitle

\section{Introduction}\label{Sec:Intro}

Quantum computation requires the mitigation of errors that occur due to faulty controls and unwanted interactions with the environment \cite{Ken_pulse, DD99}. Fault-tolerant quantum error correction is one method for mitigating these errors with the advantage that provable arbitrary quantum computation is possible given constraints on the error rates and the error locality \cite{Gottesman2010,Shor96, KLZ98, KLZScience}. 

There are many possible error correcting codes \cite{Shor96, CalderShor96, Fivequbit, Knill, BaconShor, Kitaev, Surface_codes, Color_codes, Duan2010, Topo_subsystem} and the mapping of abstract models involving qubits on a completely connected graph to a more realistic local architecture leads to a number of choices that makes analytical comparison of codes difficult. In these systems it is typical to use simulation to determine the error correcting properties \cite{Cross_thesis, Aliferis2006, Cross2009}. Although simulation of quantum systems is difficult \cite{Feynman, LloydScience}, simulation of error correction can be done efficiently for stabilizer codes where the process of error correction only includes gates in the Clifford group \cite{Gottesman_thesis, Aaronson}.  

A standard error model is a depolarizing channel where a Pauli operator from a chosen probability distribution is applied at every possible error position \cite{SteanePRA2003, Knill, Surface_codes, Cross2009}.  The depolarizing channel efficiently simulates common laboratory processes such as dephasing. It also serves as a good approximation for most error process that lead to a steady-state in which the system becomes maximally mixed. These are unital channels that map completely mixed states to completely mixed states.  

In nature it is also common to encounter interactions with the environment that lead to non-unital error channels in which the maximally mixed states are not a fixed point of the error process.  One example is amplitude damping where, given enough time, all density matrices map to a single pure state.  If an error channel is far from unital, then simulating it with Pauli errors gives large approximation errors making it hard to extract useful results.

In this paper, we go beyond simulating errors with the conventional Pauli depolarizing channel (PC). Instead of only restricting to Pauli errors, we allow any subset of efficiently simulable gate errors to occur. In particular, we look at subsets generated by including all Clifford group operators and/or Pauli measurements to the PC channel. We show that adding Clifford errors and/or measurement errors always results in more accurate approximations and results in significant improvements for most error channels. We consider an approximating error channel to be valid if it has a smaller fidelity than the target error channel and choose the best valid approximation by minimizing the Hilbert-Schmidt metric.

The paper is organized as follows. In Section \ref{sec:realchannels} we first describe the simulable error processes and introduce our expansions to the PC.  We then mention two important error channels that cannot be simulated in the stabilizer formalism and finally we describe our method for approximating a general error channel with our new models.  In Section \ref{sec:results} we compare how well these models approximate the two error channels mentioned in Section \ref{sec:realchannels} and also a collection of random error channels. In Section \ref{sec:conclusions}, we conclude and describe future research directions.

Throughout the paper we use $\sigma_1=X$, $\sigma_2=Y$, and $\sigma_3~=~Z$ to represent the Pauli matrices with associated eigenvectors $\{|+\>, |-\>\}$, $ \{|+i\>, |-i\>\}$, and $ \{|0\>, |1\>\}$ respectively.

\section{Error channels}\label{sec:realchannels}

It is convenient to consider the interaction of the environment with the system for a fixed time.  Then the system dynamics can be represented by a set of time-independent Kraus operators that form an error channel. 

We begin by considering all error channels that can be simulated efficiently within the stabilizer formalism. Next we examine two specific error channels that are outside of the stabilizer formalism.  Finally, we discuss a method by which we create an error channel that approximates a target channel. 

\subsection{Efficiently simulable error processes}

The stabilizer formalism allows for efficient simulation on a classical computer of operators from the Clifford group operating on states stabilized by Pauli operators \cite{Aaronson}. The Clifford group for $n$-qubits can be generated from CNOTs and the 1-qubit Clifford gates. As error channels, the probabilistic application of 1-qubit Clifford operators can be represented by the following Kraus operators:

\begin{list}{$\bullet$}{}
\item{Identity\\ $E_0 = \sqrt{p_0} I$}
\item{Pauli operators\\ $E_i$ = $ \sqrt{p_i} \sigma_i$ }
\item{S-like operators \\ $E_{S,\pm j}= \sqrt{p_{S,\pm j}} \exp ( -i \frac{\pi}{4}(\pm \sigma_j) )$}
\item{Hadamard-like operators\\ $E_{ j, \pm k}= \sqrt{p_{j,\pm k}} \exp (-i \frac{\pi}{2} \frac{1}{\sqrt{2}} (\sigma_j \pm \sigma_k))$ for $k > j $}
\item{Rotations about the face centers\\ $E_{ \vec{F}} = \sqrt{p_{\vec{F}}} \exp (-i \frac{\pi}{3} \sigma_{\vec{F}})$ , where $\sigma_{\vec{F}} = \vec{F} \cdot \vec{\sigma}$ and $\vec{F}$ is the unit vector from the origin to one of the eight faces of the 1-qubit Clifford octahedron \cite{vanDam2009}.}

\end{list}

The stabilizer formalism also includes non-unital operators. The simplest are measurement operations in the Pauli basis. More intricate non-unital Kraus operations can be represented as measurements followed by gates conditioned on the measurement outcomes.

We limit ourselves to non-unital operators that result in translations along the Pauli axes. For each eigenstate, $f$, of a Pauli operator,  we define the following two Kraus operators with the same classical probability:

\begin{list}{$\bullet$}{}
\item{Measurement-induced translations \\ $E_{\ket{f}\bra{f}} =\sqrt{p_{\ket{f}}} \ket{f}\bra{f} \\ E_{\ket{f}\bra{f^\perp}} = \sqrt{p_{\ket{f}}}\ket{f}\bra{f^\perp}$}
\end{list}

Notice that the effect of these two operators is to discard the state with a probability of $p_{\ket{f}}$ and replace it by $\ket{f}$. The effect on a state, when represented on the Bloch sphere, is to translate it toward $\ket{f}$. 

To ensure trace preservation, we set $p_0=\sqrt{1-\sum_a p_a}$ where $a$ sums over all other operators.

Throughout the paper we will refer to four sets of these error process: PC, CC, PMC, and CMC. The Pauli Channel (PC), introduced above, is limited to Pauli errors. The Clifford Channel (CC) includes all efficiently simulable unitary gates \footnote{During the preparation of this manuscript, we learned about independent and related work on the Clifford channel, E. Magesan, D. Puzzuoli, C.E. Granade, and D.G. Cory, arXiv:1206.5407v1 }. The Pauli and Measurement Channel (PMC)  includes all Pauli errors and all measurement-induced translation errors. Finally, the Clifford and Measurement Channel (CMC) includes all Clifford errors and all measurement-induced translation errors. We use these as approximation channels to the error channels presented below.

\begin{table}[htdp]
\caption{Kraus operators corresponding to the 4 efficiently simulable error channels.}
\begin{center}
\begin{tabular}{| c ||  l |}
\hline
\multicolumn{1}{| c ||}{ Channel Label } & \multicolumn{1}{|c|}{ Kraus error set } \\ \hline\hline
PC & $\{E_i\}$\\ \hline
PMC & $\{E_i, E_{\ket{f}\bra{f}}, E_{\ket{f}\bra{f^\perp}} \}$ \\ \hline
CC & $\{E_i, E_{S,\pm j}, E_{j, \pm k}, E_{ \vec{F}}\}$ \\ \hline
CMC & $\{E_i, E_{S,\pm j}, E_{j, \pm k}, E_{ \vec{F}}, E_{\ket{f}\bra{f}} ,E_{\ket{f}\bra{f^\perp}} \}$  \\ \hline
\end{tabular}
\end{center}
\end{table}

\subsection{Examples of non-Clifford error channels}

\subsubsection{Amplitude damping}

The amplitude damping channel ADC, represented in Equation \ref{eq:adc}, is the prototypical non-unital error channel \cite{MikeAndIke}.  The ADC describes the energy dissipation of a two-level quantum system. However small, it is present in any non-degenerate physical system.     
\begin{equation} 
\text{ADC} =
\begin{cases}
E_{A0} = |0\> \<0| \medspace + \medspace \sqrt{1-\gamma} \thinspace |1\> \<1| \\
E_{A1} = \sqrt{\gamma} \thinspace|0\> \<1|
\end{cases} 
\label{eq:adc}
\end{equation}

The rate of the energy loss to the environment, or damping, is given by the dimensionless parameter $ \gamma $, which can take any real value between 0 and 1 \footnote{$\gamma = 1 - e^{-\Gamma t}$, where $\Gamma$ is the real rate of energy loss to the environment.}.

Numerous codes have have been developed specifically to combat ADC, but studying the effects of this error channel on a circuit has yielded only a handful of results \cite{Leung97, ChuangLeung97,SingleErrorADC, Duan2010}. All of the results assume $\gamma$ to be small in order to expand the Kraus operators in a Taylor series expansion using the Pauli operator basis. 

\subsubsection{Polarization along an axis in the X-Y plane}

Another interesting error channel is a polarization along a non-Pauli axis.  Specifically, we focus on a polarization along an axis in the X-Y plane of the Bloch sphere:  
\begin{equation}
\text{Pol}_{\phi}\text{C} = 
\begin{cases}
E_{xy0} = \sqrt{1-p_\phi} \thinspace I \\
E_{xy1} = \sqrt{p_\phi} \thinspace [\cos(\phi) \thinspace X \medspace  + \medspace  \sin(\phi) \thinspace  Y]
\end{cases}
\end{equation} 
\noindent where the parameter $\phi$ represents the angle of the polarization axis with respect to the X axis and $p_\phi$ the probability of error. 

Unlike the ADC channel, $\text{Pol}_\phi \text{C}$ is unital. Yet unless the angle $\phi$ is a half-integer multiple of $\pi$, the depolarization occurs along a non-Pauli axis, and the quality of the PC approximation will vary with $\phi$.  

\subsection{Evaluating the approximations}

To study how closely our error models approximate target error channels, we compute the distance between the process matrix of our error model and the process matrix of the target error.  For an error model with $n$ operators (including the identity), this distance is a function of the $n-1$ linearly independent probabilities associated with the operators.  As a distance measure we employ the normalized Hilbert-Schmidt distance \cite{Distance_meas}.  This distance ranges from 0 (for two identical channels) to 1 (for two orthogonal channels).



\begin{equation}
D(\chi_{1},\chi_{2}) = \frac{1}{2N^2} \lVert \chi_{1} - \chi_{2} \rVert^2_{HS}
\end{equation}
\begin{equation}
\lVert A \rVert_{HS} = \sqrt{Tr(A^\dagger A)}
\end{equation}     

Throughout this paper $ N = 2 $, since we will only focus on the one-qubit errors.  After calculating the distance, we then minimize it over the $n-1$ independent variables.  

As our goal is to understand for which cases this error model would be an appropriate approximation, we want our model to be an upper bound to the error induced on the system.  Therefore, we perform the distance minimization with the constraint that the fidelity between the identity channel and our error model is not greater than the fidelity between the identity channel and the target error. This constraint ensures that our approximation will not underestimate the real target error. 
\begin{equation}
F(I,\text{Target}) \geqslant F(I,\text{Model})
\end{equation}
The fidelity can either be an average fidelity:
\begin{equation}
F_{av}(V, K) = \frac{1}{N^2} \sum_{i} | \text{Tr}(V^\dagger K_{i}) |^2
\end{equation}
where $\lbrace K_{i} \rbrace$ are the Kraus operators of the error channel $K$ and $V$ is a unitary transformation, or a worst-case fidelity:
\begin{equation}
F_{w}(V, K) = \underset{\rho \thinspace \epsilon \thinspace D}{\text{min}} \sum_{i} | \text{Tr}(V^\dagger K_{i} \rho) |^2
\end{equation}
where in this case the fidelity is minimized over all the density matrices $\rho$. The minimization was performed with Python's sequential least squares programming minimization subroutine.

We use the Hilbert-Schmidt distance for most of the analysis here due to ease of computation, but the method works for any distance measure or constraint \cite{Chuang_distance}.  In most cases, the worst case fidelity constraint would be appropriate for calculating lower bounds on error correction thresholds.  For certain cases, such as Pol$_{\phi}$C, the two constraints give the same results.

\begin{figure}[h!]
\centering
\includegraphics[scale=0.7]{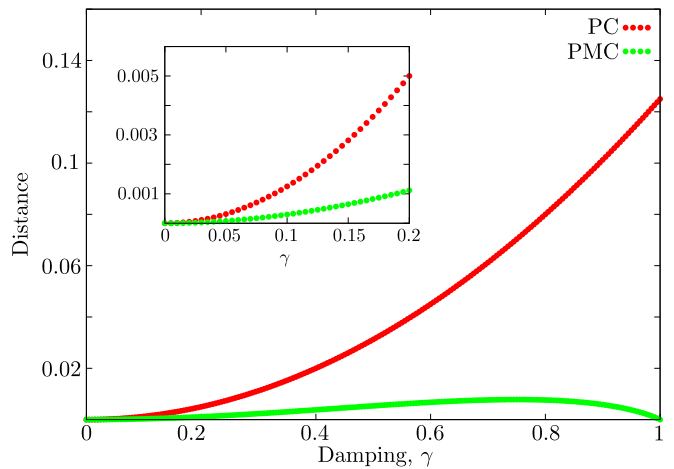}
\caption{Minimum distance between two approximate error models and the amplitude damping channel as a function of $ \gamma $, the damping strength.  Although not shown, the results for the CC and CMC are exactly the same as the results for PC and PMC, respectively.  The inset figure, a zoomed version of the same plot, gives an idea of how fast an error model without measurement-induced translations becomes an inaccurate approximation compared to an error model that includes them. For small values of $\gamma$, however, both distances scale quadratically.}
\label{fig:ADCDist}
\end{figure}

\section{Results}\label{sec:results}

\subsection{Amplitude Damping Channel (ADC)}
Figure \ref{fig:ADCDist} shows the results of the approximation of the ADC by the error models introduced in Section \ref{sec:realchannels} with the average fidelity constraint.  Each one of the 200 points corresponds to a numerical minimization for a particular damping strength. After fitting these points and then solving symbolically, for both the PC and the CC the distance between the ADC and the best approximation was found to be $D_{P} = \frac{\gamma^2}{8}$, where $\gamma$ is the damping strength.  This means that as the non-unital character of the ADC becomes more pronounced, the unital error models give less an accurate approximation.  The larger repertoire of operators in the Clifford group does not improve the approximation obtained with only Pauli operators.

On the other hand, the addition of the measurement-induced translations considerably improves the approximation.  In this case, the distance between the approximation and the ADC is given by $ D_{m} = \frac{(\gamma - 1)(\gamma + 2 \sqrt{1-\gamma} - 2 )}{8}$, and the PMC and CMC significantly outperform the models without measurement for $\gamma>0.05$. 
The PMC and CMC can match the ADC perfectly only for $\gamma = 0$, which corresponds to the trivial case, and $\gamma = 1$, which corresponds to a measurement that is actually part of the operator repertoire of our error model.  Interestingly, despite the large amount of operators in the CMC error model, the best approximation only employs the identity and the translation towards $|0 \>$ and it is given by $ \left \lbrace E_{0} = \sqrt{1-p_{m}} \thinspace I \thinspace , \thickspace E_{1} = \sqrt{p_{m}} \thinspace |0\> \<0| \thinspace , \thickspace E_{2} = \sqrt{p_{m}} \thinspace |0\> \<1| \right \rbrace$, with $p_{m} = \frac{1}{2} (1 + \gamma - \sqrt{1-\gamma})$.  It is also noteworthy that, for small $\gamma$ values, $D_{m} = \frac{\gamma^2}{32} - \frac{\gamma^3}{64} + O(\gamma^4)$, while $D_{P} = \frac{\gamma^2}{8}$: although the measurement operators improve the approximation even for small $\gamma$ values, both methods have a quadratic dependence on $\gamma$.

When the constraint is changed from the average fidelity to the worst fidelity, then the PC and CC approximations have a distance of $D_{P,w} = \frac{2\gamma^{2} - 3\gamma + 2 + 2\gamma \sqrt{1-\gamma} - 2\sqrt{1-\gamma}}{4}$, while the PMC and CMC have a distance of $D_{m,w} = 2D_{m}$.  Both of these cases result in larger distances than the ones with the average fidelity approximation and the difference between models with and without measurements is even more pronounced.

\begin{figure}[htb!]
\centering
\includegraphics[scale=0.7]{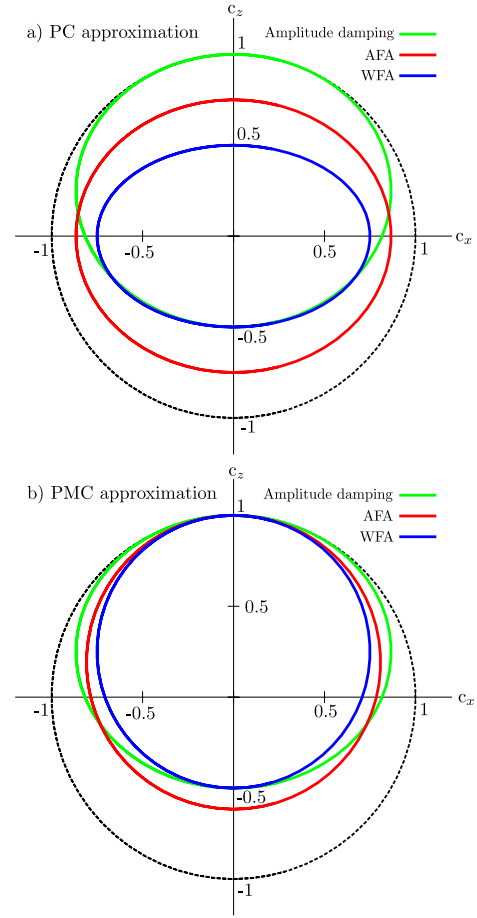}
\caption{Cross-sectional view of the Bloch sphere and the effect of amplitude damping and two approximations with different constraints.  AFA stands for average constraint approximation, while WFA stands for worst constraint approximation.  a) Channels without measurement operators.  b) Channels with measurement operators.  For both cases, $\gamma = 0.25$.}
\label{fig:ADCcircles}
\end{figure}

The results obtained by the average fidelity and the worst fidelity contraints are best illustrated in Figure \ref{fig:ADCcircles}.  Here we examine, for $\gamma = 0.25$, the closest PC (a) and PMC (b) approximation assuming either one of the two constraints. The figure shows a cross section of the Bloch sphere and its transformation by the ADC and the closest approximate channel with either the average fidelity constraint (red) or the worst fidelity constraint (blue).  Notice that for these error channels the deformed Bloch sphere is still symmetric with respect to rotations around $z$, so a cross section is enough to visualize the whole process.  

The approximation using the worst fidelity constraint guarantees that the largest distance between any input and the target channel output will be less than the largest distance between any input and the approximate channel output.  In this case, for both the ADC and its approximations the largest discrepancy between input and output occurs when the initial state is $|1 \>$.  Notice that for the PMC approximation this constraint also guarantees that for all inputs the approximate channel outputs are further from the input than for the target channel.  This is pictorially represented in Figure \ref{fig:ADCcircles}(b), where the blue curve is always \textit{inside} the green curve or further away from the initial states (black curve).  For the PC, however, this is not the case, as Figure \ref{fig:ADCcircles}(a) shows.  Here the blue curve lies \textit{outside} the green curve for some points.  Indeed, if we use a unital channel to approximate a non-unital one, it is impossible to satisfy the condition that for every input the approximate channel output will be further from the input than for the target channel.  Simply consider the maximally mixed state, which is mapped to itself by a unital channel, but mapped to a different state by a non-unital one.  

\subsection{Polarization along an axis in the X-Y plane (Pol$_{\phi}$C)} 

\begin{figure}[htb!]
\centering
\includegraphics[scale=0.7]{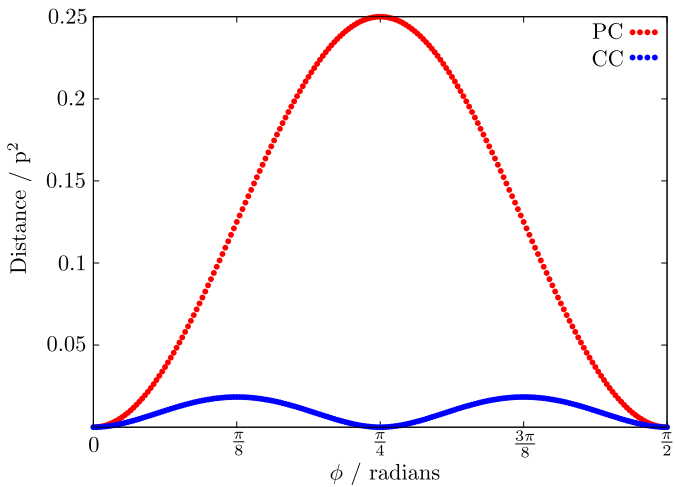}
\caption{Minimum distance between several approximate error models and the polarization along an axis in the X-Y plane of the Bloch sphere as a function of the polarization angle.  Although not shown, the results for PMC and CMC are the same as the results for PC and CC, respectively.  The distances scale quadratically with p, so the results are normalized by p$^2$.} \label{fig:xy}
\end{figure}

Figure \ref{fig:xy} shows the results of the approximation of the Pol$_{\phi}$C by the error models introduced earlier.  Once again, each one of the 200 points corresponds to a numerical minimization.  Because of the unital nature of this channel, it is the addition of the Clifford operators rather than the measurement operators that improve the approximation.  For both the PC and the PMC, the distance between Pol$_{\phi}$C and the best approximation was found to be $D_{P} = \frac{1}{4} p^2 \sin^2(2 \phi)$.  When the Clifford operators are included in the approximate channel, the new distance is reduced to $D_{C} = \frac{3}{28} p^2 (\sin(2\phi) + \cos(2\phi) -1)^2$  for $ 0 \leq \phi \leq \pi/4 $ and for $p < 0.9$ \footnote{The minimum distance is given by this function only when $p \leq \frac{7}{6+\sqrt{2}} \approx 0.944$.  Because we are interested in small errors, we will not deal with the $p > \frac{7}{6+\sqrt{2}}$ case.  Furthermore, the expression $D_{C} = \frac{3}{28} p^2 (\sin(2\phi) + \cos(2\phi) -1)^2$ is only valid for $0 < \phi < \frac{\pi}{4}$.  For other intervals, the distance is the same expression translated by the corresponding amount.  For example, for the $\frac{\pi}{4} < \phi < \frac{\pi}{2}$ interval, the distance is $D_{C} = \frac{3}{28} p^2 (\sin[2(\phi - \pi/4)] + \cos[2(\phi - \pi/4)] -1)^2$.}.  At the worst point of the CC (which in this interval occurs at $\phi = \pi/8, 3\pi/8$), the PC is 6.8 times worse.   Notice that not only the distance is decreased; the period of the distance function is also reduced from $\frac{\pi}{2}$ to $\frac{\pi}{4}$, because between every two Pauli axes there is a Clifford axis.            

Once again, despite the large amount of operators in the CMC, the best approximation uses a small number of them: the identity and the two axes closest to the polarization axis. If we only employ Pauli axes, the best approximation is $ \left \lbrace E_{0} = \sqrt{1-p} \thinspace I \thinspace , \thickspace E_{1} = \sqrt{p} \thinspace \cos(\phi) \thinspace X \thinspace , \thickspace E_{2} = \sqrt{p} \thinspace \sin(\phi) \thinspace Y \right \rbrace$, where $\phi$ and $p$ are the same as in Equation (2).  If we employ the whole Clifford group, the best approximation is given by$ \left \lbrace E_{0} = \sqrt{1-p_{1}-p_{2}} \thinspace I \thinspace , \thickspace E_{1} = \sqrt{p_{1}} \thinspace X \thinspace , \thickspace E_{2} = \sqrt{p_{2}} \thinspace H_{XY} \right \rbrace$, where $H_{XY} = \frac{1}{\sqrt{2}} (X + Y)$, $p_{1} = \frac{p}{7} (3 + 4\cos(2\phi) - 3\sin (2\phi)) $, and $p_{2} = \frac{p}{7} (3 - 3\cos{2\phi} + 4\sin(2\phi))$.  Finally, as mentioned before, for this error channel there is no difference between the results obtained with either fidelity constraint.

\subsection{Random Error Channels}

We have seen that the addition of the measurement-induced translations and the Clifford operators improves the approximation of two specific error channels.  To determine how the method works for general errors, we generated 2000 random process matrices and computed the distance of the best approximation that each one of the 4 approximate channels could make. For the 1-qubit case, a process matrix is a $4\times4$ Hermitian positive matrix $M$ with 4 constraints: Tr$(M)$ = $2$, Re$(M_{01})$ = -Im$(M_{23})$ , Re$(M_{02})$ = Im$(M_{13})$ , and Re$(M_{03})$ = -Im$(M_{12})$ \footnote{These constraints apply when the process matrix is expressed in the normalized Pauli basis}.  To generate this matrix we first create a $4\times4$ diagonal matrix $D$ with real, positive diagonal entries that add to 2.  We then create a $4\times4$ random unitary matrix $U$ and apply this unitary transformation to $D$ to obtain $M = UDU^\dagger$, which is positive with trace 2.  We then enforce the last 3 constraints mentioned earlier and keep the random process if the matrix is still positive.   

\begin{figure}[htb!]
\centering
\includegraphics[scale=0.8]{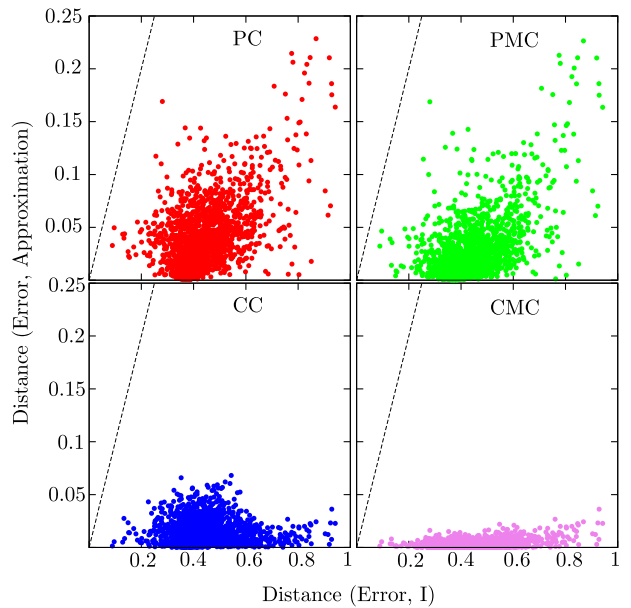}
\caption{Distance between the random error channels and the best approximations attained with each model as a function of the distance between the error and an errorless channel (identity).  The slope of a line joining the origin and a point represents the distance of the best approximation to that error relative to the magnitude of the error.  Every approximation includes the errorless channel and this limits the distance between the approximation channel and error channel to be below a line of slope 1 (black dotted line).} \label{fig:rand}
\end{figure}

\begin{table}[htdp]
\caption{Summary of the approximations obtained with each of the 4 error models.}
\begin{center}
\begin{tabular}{| c ||  c | c | c |}
\hline
\multicolumn{1}{| c ||}{ Channel} & \multicolumn{1}{| c |}{ Distance mean } & \multicolumn{1}{| c |}{ Distance median } & \multicolumn{1}{| c |}{ Distance variance } \\ \hline\hline
PC & $0.043$ & $0.038$ & $1.0 \times 10^{-3}$ \\ \hline
PMC & $0.029$ & $0.012$ & $9.1 \times 10^{-4}$ \\ \hline
CC & $0.015$ & $0.012$ & $1.5 \times 10^{-4}$ \\ \hline
CMC & $0.0027$ & $0.0011$ & $1.6 \times 10^{-5}$ \\ \hline
\end{tabular}
\end{center} \label{table:summary}
\end{table}

Figure \ref{fig:rand} illustrates the distance between each random error channel and the best approximation as a function of the distance between the error channel and the identity channel.  The fidelity constraint guarantees that the approximation will always be more distant from the identity than the error is.  Notice that as the amount of operators in the error models increases, both the mean and the median distance between each model and the random error decreases and the distributions become more compact, as summarized in Table \ref{table:summary}. 

Although the approximations with the CC had a smaller mean distance and a more compact distribution than the ones with the PMC, it is not clear that this difference is significant.  The most important improvement occurs when we add both the unital Clifford gates and the non-unital measurement-induced translations.  For the CMC, for 48$\%$ of the generated random process matrices the distance of the best approximation was less than 0.001, while for the other error channels the fraction of approximations with a distance in this interval was not greater than 6$\%$.   

\section{Conclusions} \label{sec:conclusions}

We have presented an extension to the random Pauli error model which is still compatible with efficient simulation using the Gottesman-Knill theorem and leads to a computationally tractable description of realistic error models like amplitude damping.  For ease of calculation, we have used the average distance as the measure to be optimized under two different fidelity constraints of the error channels relative to an error free channel. We have also only focused on single qubit errors in the absence of quantum operations. Once we consider simulating operations over many qubits, we will need a distance measure that is composable over tensors. A more natural distance measure in this regard is the diamond norm \cite{KitaevBook}. 

Our method can be extended to multi-qubit channels but the optimization becomes more difficult as the number of Clifford operators grows quickly with $n$.  In many cases, symmetries of the underlying error channels will minimize the number of Clifford operators that must be considered.  In future work, we will compare for a specific error correction circuit how the logical error rate compares for the models.  We expect in the case of multiple rounds of error correction a substantial difference between error models with distinct fixed points.

\begin{acknowledgements}
The authors thank Aram Harrow for valuable comments.  AV and KRB acknowledge support from the National Science Foundation through the Quantum Information for Quantum Chemistry Center for Chemical Innovation (CHE-1037992). MG, LS, and KRB were also supported by the Office of the Director of National Intelligence - Intelligence Advanced Research Projects Activity through Army Research Office contract W911NF-10-1-0231 and Department of Interior contract D11PC20167. Disclaimer: The views and conclusions contained herein are those of the authors and should not be interpreted as necessarily representing the official policies or endorsements, either expressed or implied, of IARPA, DoI/NBC, or the U.S. Government.
\end{acknowledgements}


\begin{thebibliography}{34}
\expandafter\ifx\csname natexlab\endcsname\relax\def\natexlab#1{#1}\fi
\expandafter\ifx\csname bibnamefont\endcsname\relax
  \def\bibnamefont#1{#1}\fi
\expandafter\ifx\csname bibfnamefont\endcsname\relax
  \def\bibfnamefont#1{#1}\fi
\expandafter\ifx\csname citenamefont\endcsname\relax
  \def\citenamefont#1{#1}\fi
\expandafter\ifx\csname url\endcsname\relax
  \def\url#1{\texttt{#1}}\fi
\expandafter\ifx\csname urlprefix\endcsname\relax\def\urlprefix{URL }\fi
\providecommand{\bibinfo}[2]{#2}
\providecommand{\eprint}[2][]{\url{#2}}

\bibitem[{\citenamefont{Brown et~al.}(2004)\citenamefont{Brown, Harrow, and
  Chuang}}]{Ken_pulse}
\bibinfo{author}{\bibfnamefont{K.~R.} \bibnamefont{Brown}},
  \bibinfo{author}{\bibfnamefont{A.~W.} \bibnamefont{Harrow}},
  \bibnamefont{and} \bibinfo{author}{\bibfnamefont{I.~L.}
  \bibnamefont{Chuang}}, \bibinfo{journal}{Phys. Rev. A}
  \textbf{\bibinfo{volume}{70}}, \bibinfo{pages}{052318}
  (\bibinfo{year}{2004}); \bibinfo{journal}{Phys. Rev. A}
  \textbf{\bibinfo{volume}{72}}, \bibinfo{pages}{039905(E)}
  (\bibinfo{year}{2005}).

\bibitem[{\citenamefont{Viola et~al.}(1999)\citenamefont{Viola, Knill, and
  Lloyd}}]{DD99}
\bibinfo{author}{\bibfnamefont{L.}~\bibnamefont{Viola}},
  \bibinfo{author}{\bibfnamefont{E.}~\bibnamefont{Knill}}, \bibnamefont{and}
  \bibinfo{author}{\bibfnamefont{S.}~\bibnamefont{Lloyd}},
  \bibinfo{journal}{Phys. Rev. Lett.} \textbf{\bibinfo{volume}{82}},
  \bibinfo{pages}{2417} (\bibinfo{year}{1999}).

\bibitem[{\citenamefont{Gottesman}({2010})}]{Gottesman2010}
\bibinfo{author}{\bibfnamefont{D.}~\bibnamefont{Gottesman}}, in
  \emph{\bibinfo{booktitle}{{Quantum Information Science and its Contributions
  To Mathematics}}} (\bibinfo{year}{{2010}}), vol.~\bibinfo{volume}{{68}} of
  \emph{\bibinfo{series}{{Proceedings of Symposia in Applied Mathematics}}},
  pp. \bibinfo{pages}{{13--58}}.

\bibitem[{\citenamefont{Shor}(1995)}]{Shor96}
\bibinfo{author}{\bibfnamefont{P.~W.} \bibnamefont{Shor}},
  \bibinfo{journal}{Phys. Rev. A} \textbf{\bibinfo{volume}{52}},
  \bibinfo{pages}{R2493} (\bibinfo{year}{1995}).

\bibitem[{\citenamefont{Knill et~al.}(1998{\natexlab{a}})\citenamefont{Knill,
  Laflamme, and Zurek}}]{KLZ98}
\bibinfo{author}{\bibfnamefont{E.}~\bibnamefont{Knill}},
  \bibinfo{author}{\bibfnamefont{R.}~\bibnamefont{Laflamme}}, \bibnamefont{and}
  \bibinfo{author}{\bibfnamefont{W.~H.} \bibnamefont{Zurek}},
  \bibinfo{journal}{Proceedings: Mathematical, Physical and Engineering
  Sciences} \textbf{\bibinfo{volume}{454}}, \bibinfo{pages}{pp. 365}
  (\bibinfo{year}{1998}{\natexlab{a}}).

\bibitem[{\citenamefont{Knill et~al.}(1998{\natexlab{b}})\citenamefont{Knill,
  Laflamme, and Zurek}}]{KLZScience}
\bibinfo{author}{\bibfnamefont{E.}~\bibnamefont{Knill}},
  \bibinfo{author}{\bibfnamefont{R.}~\bibnamefont{Laflamme}}, \bibnamefont{and}
  \bibinfo{author}{\bibfnamefont{W.~H.} \bibnamefont{Zurek}},
  \bibinfo{journal}{Science} \textbf{\bibinfo{volume}{279}},
  \bibinfo{pages}{342} (\bibinfo{year}{1998}{\natexlab{b}}).

\bibitem[{\citenamefont{Calderbank and Shor}(1996)}]{CalderShor96}
\bibinfo{author}{\bibfnamefont{A.~R.} \bibnamefont{Calderbank}}
  \bibnamefont{and} \bibinfo{author}{\bibfnamefont{P.~W.} \bibnamefont{Shor}},
  \bibinfo{journal}{Phys. Rev. A} \textbf{\bibinfo{volume}{54}},
  \bibinfo{pages}{1098} (\bibinfo{year}{1996}).

\bibitem[{\citenamefont{DiVincenzo and Shor}(1996)}]{Fivequbit}
\bibinfo{author}{\bibfnamefont{D.~P.} \bibnamefont{DiVincenzo}}
  \bibnamefont{and} \bibinfo{author}{\bibfnamefont{P.~W.} \bibnamefont{Shor}},
  \bibinfo{journal}{Phys. Rev. Lett.} \textbf{\bibinfo{volume}{77}},
  \bibinfo{pages}{3260} (\bibinfo{year}{1996}).

\bibitem[{\citenamefont{Knill}(2005)}]{Knill}
\bibinfo{author}{\bibfnamefont{E.}~\bibnamefont{Knill}},
  \bibinfo{journal}{Nature} \textbf{\bibinfo{volume}{434}}, \bibinfo{pages}{39}
  (\bibinfo{year}{2005}).

\bibitem[{\citenamefont{Bacon}(2006)}]{BaconShor}
\bibinfo{author}{\bibfnamefont{D.}~\bibnamefont{Bacon}},
  \bibinfo{journal}{Phys. Rev. A} \textbf{\bibinfo{volume}{73}},
  \bibinfo{pages}{012340} (\bibinfo{year}{2006}).

\bibitem[{\citenamefont{Kitaev}(2003)}]{Kitaev}
\bibinfo{author}{\bibfnamefont{A.}~\bibnamefont{Kitaev}},
  \bibinfo{journal}{Annals of Physics} \textbf{\bibinfo{volume}{303}},
  \bibinfo{pages}{2 } (\bibinfo{year}{2003}).

\bibitem[{\citenamefont{Dennis et~al.}(2002)\citenamefont{Dennis, Kitaev,
  Landahl, and Preskill}}]{Surface_codes}
\bibinfo{author}{\bibfnamefont{E.}~\bibnamefont{Dennis}},
  \bibinfo{author}{\bibfnamefont{A.}~\bibnamefont{Kitaev}},
  \bibinfo{author}{\bibfnamefont{A.}~\bibnamefont{Landahl}}, \bibnamefont{and}
  \bibinfo{author}{\bibfnamefont{J.}~\bibnamefont{Preskill}},
  \bibinfo{journal}{J. Math. Phys.} \textbf{\bibinfo{volume}{43}},
  \bibinfo{pages}{4452} (\bibinfo{year}{2002}).

\bibitem[{\citenamefont{Bombin and Martin-Delgado}(2006)}]{Color_codes}
\bibinfo{author}{\bibfnamefont{H.}~\bibnamefont{Bombin}} \bibnamefont{and}
  \bibinfo{author}{\bibfnamefont{M.~A.} \bibnamefont{Martin-Delgado}},
  \bibinfo{journal}{Phys. Rev. Lett.} \textbf{\bibinfo{volume}{97}},
  \bibinfo{pages}{180501} (\bibinfo{year}{2006}).

\bibitem[{\citenamefont{Duan et~al.}({2010})\citenamefont{Duan, Grassl, Ji, and
  Zeng}}]{Duan2010}
\bibinfo{author}{\bibfnamefont{R.}~\bibnamefont{Duan}},
  \bibinfo{author}{\bibfnamefont{M.}~\bibnamefont{Grassl}},
  \bibinfo{author}{\bibfnamefont{Z.}~\bibnamefont{Ji}}, \bibnamefont{and}
  \bibinfo{author}{\bibfnamefont{B.}~\bibnamefont{Zeng}}, in
  \emph{\bibinfo{booktitle}{{2010 IEEE International Symposium On Information
  Theory}}} (\bibinfo{year}{{2010}}), {IEEE International Symposium on
  Information Theory}, pp. \bibinfo{pages}{{2672--2676}}.

\bibitem[{\citenamefont{Bombin}(2010)}]{Topo_subsystem}
\bibinfo{author}{\bibfnamefont{H.}~\bibnamefont{Bombin}},
  \bibinfo{journal}{Phys. Rev. A} \textbf{\bibinfo{volume}{81}},
  \bibinfo{pages}{032301} (\bibinfo{year}{2010}).

\bibitem[{\citenamefont{Cross}(2009)}]{Cross_thesis}
\bibinfo{author}{\bibfnamefont{A.}~\bibnamefont{Cross}}, Ph.D. thesis,
  \bibinfo{school}{Massachusetts Institute of Technology}
  (\bibinfo{year}{2009}).

\bibitem[{\citenamefont{Aliferis et~al.}(2006)\citenamefont{Aliferis,
  Gottesman, and Preskill}}]{Aliferis2006}
\bibinfo{author}{\bibfnamefont{P.}~\bibnamefont{Aliferis}},
  \bibinfo{author}{\bibfnamefont{D.}~\bibnamefont{Gottesman}},
  \bibnamefont{and} \bibinfo{author}{\bibfnamefont{J.}~\bibnamefont{Preskill}},
  \bibinfo{journal}{Quantum Info. Comput.} \textbf{\bibinfo{volume}{6}},
  \bibinfo{pages}{97} (\bibinfo{year}{2006}), ISSN \bibinfo{issn}{1533-7146}.

\bibitem[{\citenamefont{Cross et~al.}(2009)\citenamefont{Cross, Divincenzo, and
  Terhal}}]{Cross2009}
\bibinfo{author}{\bibfnamefont{A.~W.} \bibnamefont{Cross}},
  \bibinfo{author}{\bibfnamefont{D.~P.} \bibnamefont{Divincenzo}},
  \bibnamefont{and} \bibinfo{author}{\bibfnamefont{B.~M.}
  \bibnamefont{Terhal}}, \bibinfo{journal}{Quantum Info. Comput.}
  \textbf{\bibinfo{volume}{9}}, \bibinfo{pages}{541} (\bibinfo{year}{2009}),
  ISSN \bibinfo{issn}{1533-7146}.

\bibitem[{\citenamefont{Feynman}(1082)}]{Feynman}
\bibinfo{author}{\bibfnamefont{R.}~\bibnamefont{Feynman}},
  \bibinfo{journal}{International Journal of Theoretical Physics}
  \textbf{\bibinfo{volume}{21}}, \bibinfo{pages}{467} (\bibinfo{year}{1082}).

\bibitem[{\citenamefont{Lloyd}(1996)}]{LloydScience}
\bibinfo{author}{\bibfnamefont{S.}~\bibnamefont{Lloyd}},
  \bibinfo{journal}{Science} \textbf{\bibinfo{volume}{273}},
  \bibinfo{pages}{1073} (\bibinfo{year}{1996}).

\bibitem[{\citenamefont{Gottesman}(1997)}]{Gottesman_thesis}
\bibinfo{author}{\bibfnamefont{D.}~\bibnamefont{Gottesman}}, Ph.D. thesis,
  \bibinfo{school}{California Institute of Technology} (\bibinfo{year}{1997}).

\bibitem[{\citenamefont{Aaronson and Gottesman}(2004)}]{Aaronson}
\bibinfo{author}{\bibfnamefont{S.}~\bibnamefont{Aaronson}} \bibnamefont{and}
  \bibinfo{author}{\bibfnamefont{D.}~\bibnamefont{Gottesman}},
  \bibinfo{journal}{Phys. Rev. A} \textbf{\bibinfo{volume}{70}},
  \bibinfo{pages}{052328} (\bibinfo{year}{2004}).

\bibitem[{\citenamefont{Steane}(2003)}]{SteanePRA2003}
\bibinfo{author}{\bibfnamefont{A.~M.} \bibnamefont{Steane}},
  \bibinfo{journal}{Phys. Rev. A} \textbf{\bibinfo{volume}{68}},
  \bibinfo{pages}{042322} (\bibinfo{year}{2003}).


\bibitem[{\citenamefont{vanDam}(2009)}]{vanDam2009}
\bibinfo{author}{\bibfnamefont{W.}~\bibnamefont{van}~\bibnamefont{Dam}} \bibnamefont{and} 
\bibinfo{author}{\bibfnamefont{M.}~\bibnamefont{Howard}},
  \bibinfo{journal}{Phys. Rev. Lett.} \textbf{\bibinfo{volume}{103}},
  \bibinfo{pages}{170504} (\bibinfo{year}{2009}).

\bibitem[{Not({\natexlab{a}})}]{Note1}
\bibinfo{note}{During the preparation of this manuscript, we learned about
  independent and related work on the Clifford channel, E. Magesan, D.
  Puzzuoli, C.E. Granade, and D.G. Cory, arXiv:1206.5407v1}.

\bibitem[{\citenamefont{Nielsen and Chuang}(2001)}]{MikeAndIke}
\bibinfo{author}{\bibfnamefont{M.~A.} \bibnamefont{Nielsen}} \bibnamefont{and}
  \bibinfo{author}{\bibfnamefont{I.~L.} \bibnamefont{Chuang}},
  \emph{\bibinfo{title}{Quantum Computation and Quantum Information}}
  (\bibinfo{publisher}{Cambridge University Press},
  \bibinfo{address}{Cambridge, UK}, \bibinfo{year}{2001}).

\bibitem[{Not({\natexlab{b}})}]{Note2}
\bibinfo{note}{$\gamma = 1 - e^{-\Gamma t}$, where $\Gamma $ is the real rate
  of energy loss to the environment.}

\bibitem[{\citenamefont{Leung et~al.}(1997)\citenamefont{Leung, Nielsen,
  Chuang, and Yamamoto}}]{Leung97}
\bibinfo{author}{\bibfnamefont{D.~W.} \bibnamefont{Leung}},
  \bibinfo{author}{\bibfnamefont{M.~A.} \bibnamefont{Nielsen}},
  \bibinfo{author}{\bibfnamefont{I.~L.} \bibnamefont{Chuang}},
  \bibnamefont{and} \bibinfo{author}{\bibfnamefont{Y.}~\bibnamefont{Yamamoto}},
  \bibinfo{journal}{Phys. Rev. A} \textbf{\bibinfo{volume}{56}},
  \bibinfo{pages}{2567} (\bibinfo{year}{1997}).

\bibitem[{\citenamefont{Chuang et~al.}(1997)\citenamefont{Chuang, Leung, and
  Yamamoto}}]{ChuangLeung97}
\bibinfo{author}{\bibfnamefont{I.~L.} \bibnamefont{Chuang}},
  \bibinfo{author}{\bibfnamefont{D.~W.} \bibnamefont{Leung}}, \bibnamefont{and}
  \bibinfo{author}{\bibfnamefont{Y.}~\bibnamefont{Yamamoto}},
  \bibinfo{journal}{Phys. Rev. A} \textbf{\bibinfo{volume}{56}},
  \bibinfo{pages}{1114} (\bibinfo{year}{1997}).

\bibitem[{\citenamefont{Shor et~al.}(2011)\citenamefont{Shor, Smith, Smolin,
  and Zeng}}]{SingleErrorADC}
\bibinfo{author}{\bibfnamefont{P.~W.} \bibnamefont{Shor}},
  \bibinfo{author}{\bibfnamefont{G.}~\bibnamefont{Smith}},
  \bibinfo{author}{\bibfnamefont{J.~A.} \bibnamefont{Smolin}},
  \bibnamefont{and} \bibinfo{author}{\bibfnamefont{B.}~\bibnamefont{Zeng}},
  \bibinfo{journal}{IEEE Transactions On Information Theory}
  \textbf{\bibinfo{volume}{57}}, \bibinfo{pages}{7180} (\bibinfo{year}{2011}).

\bibitem[{\citenamefont{Grace et~al.}(2010)\citenamefont{Grace, Dominy, Kosut,
  Brif, and Rabitz}}]{Distance_meas}
\bibinfo{author}{\bibfnamefont{M.~D.} \bibnamefont{Grace}},
  \bibinfo{author}{\bibfnamefont{J.}~\bibnamefont{Dominy}},
  \bibinfo{author}{\bibfnamefont{R.~L.} \bibnamefont{Kosut}},
  \bibinfo{author}{\bibfnamefont{C.}~\bibnamefont{Brif}}, \bibnamefont{and}
  \bibinfo{author}{\bibfnamefont{H.}~\bibnamefont{Rabitz}},
  \bibinfo{journal}{New Journal of Physics} \textbf{\bibinfo{volume}{12}},
  \bibinfo{pages}{015001} (\bibinfo{year}{2010}).

\bibitem[{\citenamefont{Gilchrist et~al.}(2005)\citenamefont{Gilchrist,
  Langford, and Nielsen}}]{Chuang_distance}
\bibinfo{author}{\bibfnamefont{A.}~\bibnamefont{Gilchrist}},
  \bibinfo{author}{\bibfnamefont{N.~K.} \bibnamefont{Langford}},
  \bibnamefont{and} \bibinfo{author}{\bibfnamefont{M.~A.}
  \bibnamefont{Nielsen}}, \bibinfo{journal}{Phys. Rev. A}
  \textbf{\bibinfo{volume}{71}}, \bibinfo{pages}{062310}
  (\bibinfo{year}{2005}).

\bibitem[{Not({\natexlab{c}})}]{Note3}
\bibinfo{note}{The minimum distance is given by this function only when $p \leq
  \protect \frac {7}{6+\protect \sqrt {2}} \approx 0.944$. Because we are
  interested in small errors, we will not deal with the $p > \protect \frac
  {7}{6+\protect \sqrt {2}}$ case. Furthermore, the expression $D_{C} =
  \protect \frac {3}{28} p^2 (\protect \qopname \relax o{sin}(2\phi ) +
  \protect \qopname \relax o{cos}(2\phi ) -1)^2$ is only valid for $0 < \phi <
  \protect \frac {\pi }{4}$. For other intervals, the distance is the same
  expression translated by the corresponding amount. For example, for the
  $\protect \frac {\pi }{4} < \phi < \protect \frac {\pi }{2}$ interval, the
  distance is $D_{C} = \protect \frac {3}{28} p^2 (\protect \qopname \relax
  o{sin}[2(\phi - \pi /4)] + \protect \qopname \relax o{cos}[2(\phi - \pi /4)]
  -1)^2$.}

\bibitem[{Not({\natexlab{d}})}]{Note4}
\bibinfo{note}{These constraints apply when the process matrix is expressed in
  the normalized Pauli basis}.

\bibitem[{\citenamefont{Kitaev et~al.}(2002)\citenamefont{Kitaev, Shen, and
  Vyalyi}}]{KitaevBook}
\bibinfo{author}{\bibfnamefont{A.~Y.} \bibnamefont{Kitaev}},
  \bibinfo{author}{\bibfnamefont{A.~H.} \bibnamefont{Shen}}, \bibnamefont{and}
  \bibinfo{author}{\bibfnamefont{M.~N.} \bibnamefont{Vyalyi}},
  \emph{\bibinfo{title}{Classical and Quantum Computation}}
  (\bibinfo{publisher}{American Mathematical Society},
  \bibinfo{address}{Boston, MA, USA}, \bibinfo{year}{2002}).

\end{thebibliography}

\end{document}